\documentstyle[aps,prb,graphicx,multicol]{revtex}

\makeatletter

\renewenvironment{figure}
  {\let\@capwidth\linewidth\def\@captype{figure}}
  {}
\makeatother

\begin{document}
\draft
\preprint{arch-ive/9808144}

\title{Multiple Histogram Method for Quantum Monte Carlo}

\author{C. L. Martin\thanks{email: chris@physics.ucsb.edu}}
\address{Physics Department, University of California, Santa Barbara, CA 93106}

\date{\today}
\maketitle

\begin{abstract}
  An extension to the multiple-histogram method (sometimes
  referred to as the Ferrenberg-Swendsen method) for use in quantum
  Monte Carlo simulations is presented.  This method is shown to work
  well for the 2D repulsive Hubbard model, allowing measurements to be
  taken over a continuous region of parameters.  The method also
  reduces the error bars over the range of parameter values due the
  overlapping of multiple histograms.  A continuous sweep of
  parameters and reduced error bars allow one to make more difficult
  measurements, such as Maxwell constructions used to study phase
  separation.  Possibilities also exist for this method to be used for
  other quantum systems.
\end{abstract}


\begin{multicols}{2}

\section{Introduction}

When making calculations using the Monte Carlo method, one often would
like to make measurements of some observable as a function of the
parameters of the hamiltonian.  To do this the standard procedure is
to perform a run at one setting of the parameters until a measurement
of the observable with sufficiently small error bars is produced.  One
then moves on to another setting of the parameters, and so on until a
large discrete set of measurements of the observable is produced.
This can require large amounts of computer time depending on the
details of the hamiltonian and the Monte Carlo method one has chosen.
Building an approximation of a continuous function for the desired
observable point by point would require an even larger set of discrete
points, and a related increase in computer time.

Continuous functions of observables are useful for innumerable
activities, such as verifying functional dependencies and looking for
phase separation using Maxwell constructions.\cite{1987:Huang} The
multiple histogram method\cite{1989:Ferre.Swend} (MHM) allows one to
produce these continuous functions for a classical hamiltonian.  In
this paper, it will be shown that under certain circumstances the MHM
can also be applied to quantum mechanical hamiltonians as well, for
example the two-dimensional Hubbard model.\cite{1963:Hubba}
Observables of this hamiltonian will be measured using standard
quantum Monte Carlo (QMC) techniques\cite{qmc} and the MHM will be
applied to obtain observables as continuous functions of the
parameters of the system.  As in the classical case, the use of
overlapping histograms reduces the errors below those of a single
measurement.  Furthermore, the MHM will be applied specifically to the
density and energy of the Hubbard model as functions of the chemical
potential to search for signs of phase separation.

\section{Single Histogram Method}

In order to obtain a variable as a continuous function of
thermodynamic variables like the inverse temperature, $\beta = (k_B
T)^{-1}$, and the chemical potential, $\mu$, one can use a technique
originally used by Salsburg {\it
  et~al.},\cite{1959:Salsb.Jacob.Ficke.Wood} extended by Valleau and
Card,\cite{1972:Valle.Card} and further extended by Swendsen and
Ferrenberg.\cite{1989:Ferre.Swend,ferreandswend} To see how this can
be done for the chemical potential, $\mu$, let us assume we have a
classical hamiltonian,
\begin{equation}
  \label{swend.classical}
  {\mathcal H} = E - \mu n ,
\end{equation}
where $E$ is the energy and $n$ is the
particle density.
From this hamiltonian, one can derive a partition function for a
particular value of $\mu$,
\begin{equation}
  \label{swend.classical.partition}
  Z_{\mu} = \sum_{n} \rho(n) e^{\beta\mu n},
\end{equation}
where $\rho(n)$ is a density of states
which for these purposes we can leave undetermined.
We can also derive a probability to have a given density for
this value of $\mu$,
\begin{equation}
  \label{swend.classical.prob}
  P_{\mu}(n) = \frac{\rho(n) e^{\beta \mu n}}
  {Z_{\mu}}.
\end{equation}
Substituting Eqn.~\ref{swend.classical.partition} into
Eqn.~\ref{swend.classical.prob} allows us to find the probability for
a value of $n$ at a different value of the chemical potential, $\mu'$,
\begin{equation}
  \label{swendsen.classical.answer}
  P_{\mu'}(n) = \frac{P_{\mu}(n) 
    e^{\beta (\mu'-\mu) n}}
  {\sum_{n} P_{\mu}(n) 
    e^{\beta (\mu'-\mu) n}} .
\end{equation}
Eqn.~\ref{swendsen.classical.answer} can be extended for use with many
histograms taken at varying parameters, as discussed by Ferrenberg and
Swendsen.\cite{1989:Ferre.Swend}  Using multiple
histograms allows one to increase the range of validity of the
technique.  

\begin{figure}
  \begin{center}
    \includegraphics[angle=-90,width=\linewidth]{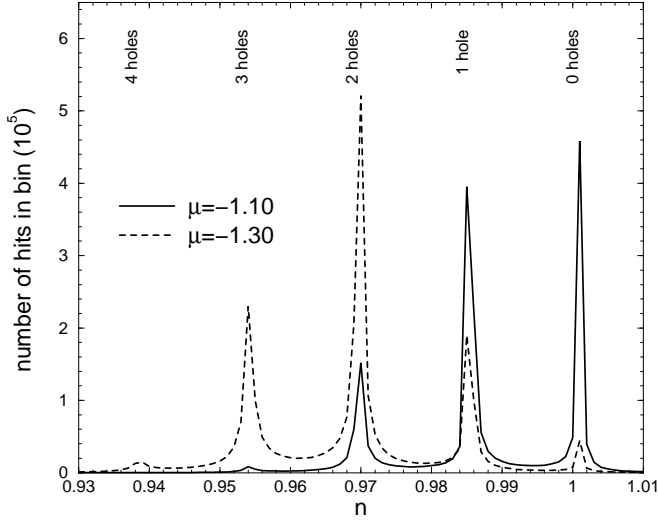}
    \caption{A histogram of the particle density sampled after each
    sweep in a Quantum Monte Carlo simulation of the Hubbard model on
    a 8 by 8 lattice with $\beta t=8$, $U/t=8$, and $\mu=-1.10$ and $-1.30$.  One
    can see well defined peaks at particle densities corresponding
    with integer numbers of holes.  It is also possible to see how the
    two histograms at different parameters overlap each other,
    allowing one to be normalized with respect to the other.
}
    \label{fig:b8.hist}
  \end{center}
\end{figure}

\begin{figure}
  \begin{center}
    \includegraphics[angle=-90,width=\linewidth]{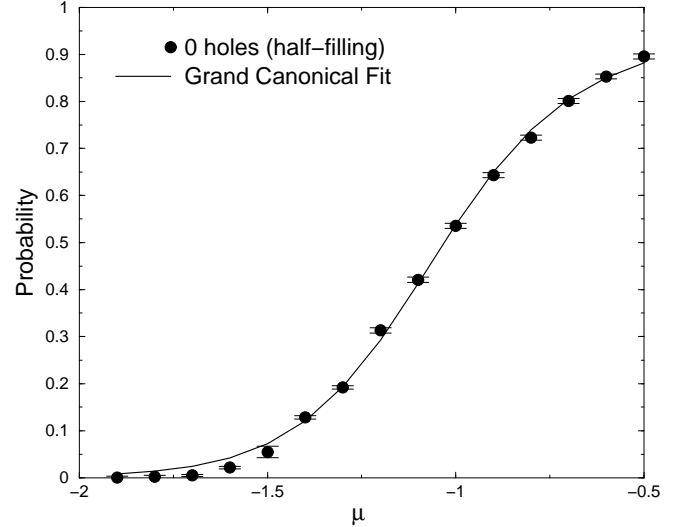}
    \caption{Normalizing the peaks in histograms such as the ones shown
    in Fig.~\protect\ref{fig:b8.hist}, yield the probability for a
    certain particle density.  One can then change the $\mu$ used in
    the simulation and follow the relative probability of this peak as
    a function of $\mu$.  If the states truly have a definite particle
    density they should follow a curve given by the grand canonical
    distribution which
    is shown in the solid line.
}
    \label{fig:ebetamun}
  \end{center}
\end{figure}

To demonstrate the use of this method for quantum hamiltonians, the Hubbard hamiltonian\cite{1963:Hubba}
will be used
\begin{eqnarray}
  \label{hubbard.hamiltonian}
H & = & -t\sum_{\langle ij\rangle} \sum_{\sigma=\{\uparrow\downarrow\}} 
\left( \hat{c}^\dagger_{i\sigma} \hat{c}_{j\sigma} + 
  \hat{c}^\dagger_{j\sigma} \hat{c}_{i\sigma} \right) \nonumber \\
&& + U \sum_i \left(\hat{n}_{i\uparrow} - \frac{1}{2}\right)
\left(\hat{n}_{i\downarrow}  - \frac{1}{2}\right)\\
&& - \mu \sum_i \left( \hat{n}_{i\uparrow} + \hat{n}_{i\downarrow}\right), \nonumber
\end{eqnarray}
where $t$ is a parameter to set the strength of electron hopping
(kinetic energy), $\hat{c}^{(\dagger)}_{i\sigma}$ is the annihilation
(creation) operator for an electron at site $i$ with spin $\sigma =
\left\{\uparrow \downarrow\right\}$, $U$ is a parameter to set the
strength of on-site coulomb repulsion, $\mu$ is the chemical
potential, and $\hat{n}_{i\sigma} = \hat{c}^\dagger_{i\sigma}
\hat{c}_{i\sigma}$.  Clearly there are quantum operators in the
Hubbard hamiltonian, so the classical derivation given earlier for the
probability as a function of $\mu$ will only hold if the system
happens to be in states of definite particle number.  In determinantal
QMC, a Hubbard-Stratonovich (HS) transformation\cite{1983:Hirsc} leads
to a bilinear fermion form for the interaction.  One sums over all
possible fermion states and then over the HS field.  Thus a given HS
configuration comes from a trace over all particle numbers so that it
is not guaranteed that a given HS configuration will have a definite
particle number.  It is this feature that distinguishes this problem
from the classical statistical mechanics problem.  However if one
finds that the various HS configurations are indeed characterized by
an integer fermion occupation, then as we will discuss, one can
proceed.  Here for the 2D Hubbard model near half-filling, the charge
gap provides an energy barrier to non-integer filling.  As can be seen
in Fig.~\ref{fig:b8.hist}, when one runs simulations at a sufficiently
low temperature, one observes peaks in the histogram corresponding to
states with integer numbers of holes.  If these states truly have a
definite number of particles, the peaks should scale using the grand
canonical distribution for runs at different values of $\mu$.  In
Fig.~\ref{fig:ebetamun}, such a test is performed and the peaks do
scale with the grand canonical distribution as desired.

Note that in Fig.~\ref{fig:b8.hist}, multiple histograms are
presented in order to emphasize the fact that the MHM allows one to
obtain even more information when several histograms are used.  By
normalizing each histogram with respect to the others, a continuous
function of the desired observable may be obtained.  In the results
that follow, we have used many overlapping histograms in
order to fully cover the range of simulation variables of interest.

In parameter regimes\cite{largeu_scalettar}
 where these peaks are present in the histograms
of particle number, one can proceed to take any operator that conserves
particle density and determine its behavior as a continuous function
of $\mu$.  This can be done by using a reduced number of simulations at
different values of $\mu$, as long as the respective histograms of
particle density overlap.  To prevent unwanted overlap from very
distant histograms, it was found necessary to introduce a Gaussian
cutoff.  The cutoff reduces the effective weight of a histogram as the
difference between the value of $\mu$ for that histogram and
the value of $\mu'$ at which the MHM method is run increases.
As an added
benefit, the error bars for this continuous function of $\mu$ will be
reduced from those obtained from any one simulation, because one can use
information about the operator from all of the simulations performed.

\begin{figure}
  \begin{center}
    \includegraphics[height=\linewidth,angle=-90]{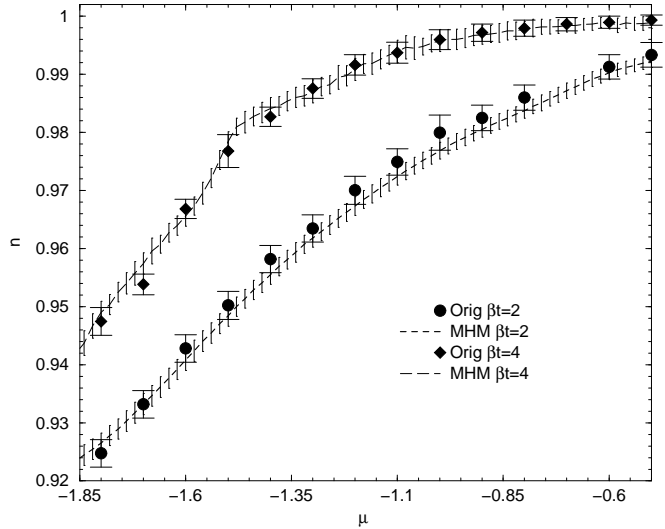}
    \caption{The original points and the MHM method continuous curves
      are shown for the particle density ($n$) vs. chemical potential
      ($\mu$) for two values of the inverse temperature, $\beta$.
      All simulations are performed on an 8 by 8 square lattice at
      $U/t = 8$.  Error bars on the original data
      points are shown as bars with hats originating from the points,
      and error bars on the MHM line are shown as bars without hats
      originating at regular intervals from the dashed line.
}
    \label{fig:nresults}
  \end{center}
\end{figure}

\section{Results}

In Fig.~\ref{fig:nresults} one can see what happens when the
Multiple Histogram Method (MHM) is applied to a set of histograms measuring the
particle number during a QMC simulation of the Hubbard model.  The
chemical potential, $\mu$, is varied over a range starting from
half-filling ($\mu = 0$ and thus $n = \langle \hat{n} \rangle = 1$)
through increasingly negative values of $\mu$.  The individual runs are
shown as points with associated error bars, and the MHM is applied to
get a continuous function of $n$ vs. $\mu$ with
error bars shown as vertical bars around the dashed line.  The error
bars are checked through both the bootstrap
method\cite{1983:Efron.Gong} and by incorporating the errors due to
correlations in the reweighted samples and their finite
size.\cite{1998:Newma.Palme}  To see how the line changes with
different values of the inverse temperature, $\beta$, the process is
repeated at $\beta t = 4$.  

As one can see, the error bars on the
MHM line are comparable to those from the original points, and the MHM
continuously fits the original points.
As the simulation temperature decreases (large $\beta$) the
sign problem\cite{1990:Loh.Guber.Scale.White} becomes much more
severe.  For the histogram method this means that while there are
peaks in the histograms for $n$ vs. $\mu$ for configurations with
positive signs, they are wiped out by nearly identical peaks with
negative signs.  

\begin{figure}
  \begin{center}
    \includegraphics[height=\linewidth,angle=-90]{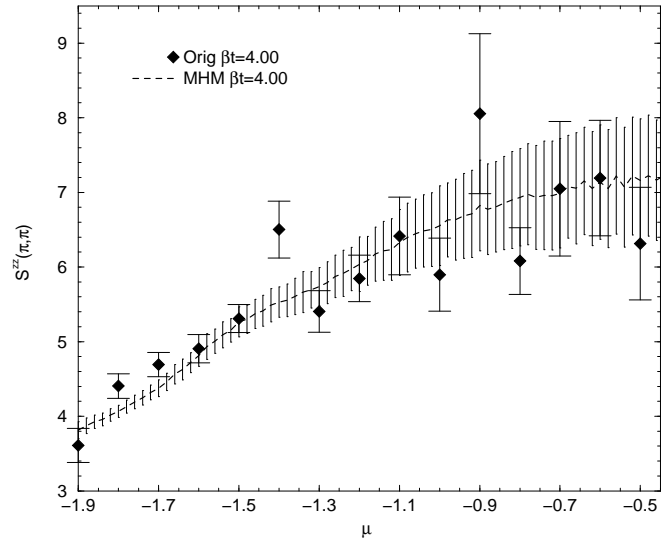}
    \caption{Original points and the resulting MHM method
      continuous curve are shown for the equal time $zz$
      antiferromagnetic spin correlation function ($S^{zz}(\pi,\pi)$).
      As the parameters are moved closer to half-filling ($\mu=0$),
      the system begins to develop antiferromagnetic order as shown by
      the increasing value of $S^{zz}(\pi,\pi)$.  The simulations are
      performed on an 8 by 8 square lattice at $U/t = 8$.  
      Error bars on the original data
      points are shown as bars with hats originating from the points,
      and error bars on the MHM line are shown as bars without hats
      originating at regular intervals from the dashed line.
      }
    \label{fig:AFZZresults}
  \end{center}
\end{figure}

In Fig.~\ref{fig:AFZZresults}, the MHM is applied to another
interesting observable, the equal time antiferromagnetic spin
correlation function, $S^{zz}(\pi,\pi)$, the Fourier transform of
$\langle (n_{i\uparrow} - n_{i\downarrow}) (n_{j\uparrow} -
n_{j\downarrow}) \rangle$ taken at momentum $(\pi,\pi)$.  As the
system moves closer to half-filling ($\mu = 0$), the antiferromagnetic
order increases as expected.\cite{1989:Hirsc.Tang}

\section{Phase Separation}

\begin{figure}
  \begin{center}
    \includegraphics[height=\linewidth,angle=-90]{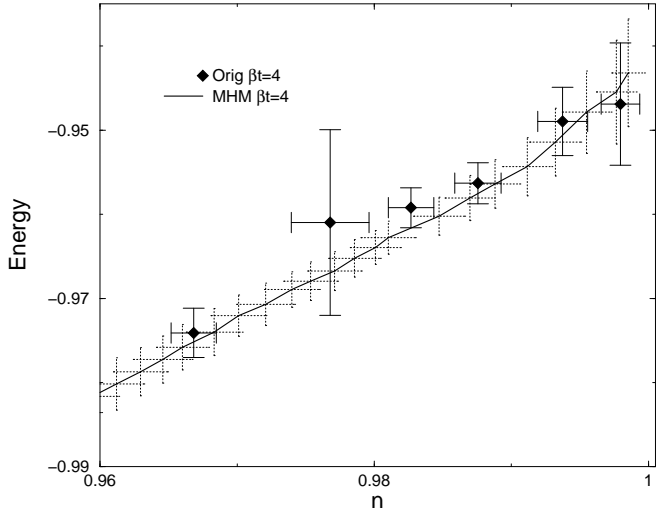}
    \caption{Original points and the resulting MHM method continuous
      curve are shown for a Maxwell construction of energy vs.
      particle density ($n$); some points have been omitted for
      clarity.  Phase separation would be indicated by a region of the
      curve with upwards curvature, which appears in neither curve.
      Error bars for both the uncertainty in energy
      and particle density on the original data points are shown as
      bars with hats originating from the points, and error bars on
      the MHM line are shown as dotted bars without hats originating
      from the dashed line.}
    \label{fig:maxwell_constuction_beta4}
  \end{center}
\end{figure}

Recently, computational evidence for the existence of phase separation
in the strongly correlated t-J and Hubbard models, both for and
against, have been presented in the literature.\cite{multicomb} One of
the best ways to search for phase separation is through the use of a
Maxwell construction.\cite{1987:Huang} A Maxwell construction consists
in its simplest form as a graph of energy vs. density.  Thermodynamics
wants a system to minimize its free energy, and so a phase separated
region will appear on the graph as a region of upwards curvature
($\frac{\partial^2 E}{\partial n^2} < 0$).  In such a region the
energy of the phase separated state will be lower than that of the
homogeneous state, and phase separation is thus favored.  It is of
interest, therefore, to apply the MHM to obtain continuous energy
versus density measurements in order to investigate the possible
existence of phase separation.  In
Fig.~\ref{fig:maxwell_constuction_beta4}, a Maxwell construction for
the 2D Hubbard model at $\beta t = 4$ is shown.  With just the
original data points it is nearly impossible to say anything
definitive about the curvature of the line, however when the MHM
method is employed, the lack of any clear signs of upwards curvature
becomes much more apparent.  Thus it is possible to conclude that one
does not see phase separation for this range of parameters, a
conclusion that would have been difficult to make without using the
full information from the histograms.

\section{Conclusions}

In some circumstances, the MHM method can be used effectively for
hamiltonians containing quantum operators, such as the Hubbard
hamiltonian just described.  
The ability
to generate continuous functions of an observable makes a number of
analysis techniques much more feasible.  For example, in order to
perform a Maxwell construction,\cite{1987:Huang} a continuous function
of energy as a function of particle density is desired.  Previously a
function like this would have been constructed laboriously one point
at a time using the averages of large simulations, and conclusions
drawn using reasonable guesses about what function would fit the
calculated points.  Using the multiple histogram method, the large
quantity of information present in the histograms of each simulation
is brought to bear using simple statistical physics arguments.  The
information from the simulations is thus used more efficiently and
smaller error bars and reasonable continuous functions are produced.

\section{Acknowledgements}

Support was received from the US Department of Energy under Grant
No. DE--FG03--85ER45197, and computer time on the Cray
T90 at SDSC was provided by NPACI.
Useful discussions and insights were generously provided by D. Scalapino,
R. Sugar, D. Duffy, and E. Kim.

\bibliography{chris,swendsen}
\bibliographystyle{prsty_chris}

\end{multicols}

\end{document}